\documentclass[aps, prl, twocolumn, lengthcheck, superscriptaddress, nofootinbib]{revtex4-1}

\usepackage{amsfonts}
\usepackage{amsmath}
\usepackage{amssymb}
\usepackage{xcolor}
\usepackage{graphicx}
\usepackage{subfigure}
\usepackage{slashed}
\usepackage{hyperref}
\usepackage{amsthm}
\usepackage{mathrsfs}
\usepackage{color}
\usepackage{epsfig}
\usepackage{epstopdf}
\usepackage{bm}
\usepackage[T1]{fontenc}
\usepackage{ae,aecompl}	
\usepackage[toc,page]{appendix}
\usepackage{soul}
\usepackage[utf8]{inputenc}
\usepackage{multirow}
\usepackage{verbatim}
\usepackage{threeparttable}

\def\be{\begin{eqnarray}}
\def\ee{\end{eqnarray}}

\def\lsim{\mathrel{\rlap{\lower3pt\hbox{\hskip1pt$\sim$}}
		\raise1pt\hbox{$<$}}} 
\def\gsim{\mathrel{\rlap{\lower3pt\hbox{\hskip1pt$\sim$}}
		\raise1pt\hbox{$>$}}} 
\newcommand\sect[1]{{\it{#1.}}}

\begin{document}
\title{Peak of sound velocity, scale symmetry, and nuclear force in baryonic matter}

\author{Lu-Qi Zhang}
\affiliation{School of Physics, Nanjing University, Nanjing 210093, China}
\affiliation{School of Frontier Sciences, Nanjing University, Suzhou 215163, China}
\affiliation{School of Fundamental Physics and Mathematical Sciences, Hangzhou Institute for Advanced Study, UCAS, Hangzhou 310024, China}

\author{Yao Ma}
\email{mayao@ucas.ac.cn}
\affiliation{School of Frontier Sciences, Nanjing University, Suzhou 215163, China}
\affiliation{School of Fundamental Physics and Mathematical Sciences, Hangzhou Institute for Advanced Study, UCAS, Hangzhou 310024, China}

\author{Yong-Liang Ma}
\email{ylma@nju.edu.cn}
\affiliation{School of Frontier Sciences, Nanjing University, Suzhou 215163, China}
\affiliation{International Center for Theoretical Physics Asia-Pacific (ICTP-AP) , UCAS, Beijing 100190, China}

\begin{abstract}

The sound velocity in homogeneous matter has fundamental significance as it relates to the stiffness of the equation of state of compact star matter. In this work, we investigate the density evolution of the sound velocity in homogeneous {neutron matter at zero temperature} by using an effective field theory implemented with a conformal compensator---the nonlinear realization of scale symmetry---regarded as the source of the lightest scalar meson. We find that the peak of sound velocity emerges naturally in the intermediate density region, $(1-2.5)n_0$, without resorting to any transitions from hadron to exotic configurations or introducing new degrees of freedom. This phenomenon is not found in the Walecka-type models where the sigma meson is included in the linear-type approach, therefore it is an intrinsic character of the dilaton compensator approach through the matching of the QCD trace anomaly; a mechanism has not been found before, and it connects to the character of the lightest scalar meson. In addition, these observations shed light on how the hidden scale symmetry manifests in the nuclear medium from the unitarity limit in dilute matter to the dilaton limit in compact star matter.

\end{abstract}

\maketitle


\sect{Introduction}
The observations of massive neutron stars with masses $\gsim 2.0 M_\odot$~\cite{Demorest:2010bx,Fonseca:2016tux,NANOGrav:2017wvv,Antoniadis:2013pzd,NANOGrav:2019jur,Fonseca:2021wxt,Romani:2022jhd} and the detection of gravitational waves emitted from binary systems including a neutron star~\cite{LIGOScientific:2017vwq,LIGOScientific:2020aai,LIGOScientific:2020zkf,LIGOScientific:2021qlt} put more stringent constraints on the equation of state (EoS) of compact star matter~\cite{Annala:2017llu,Tsang:2018kqj,Sorensen:2023zkk}. However, there is still great uncertainty about the EoS at compact star densities (see, e.g., Refs.~\cite{Chen:2015gba,Gil:2018yah,Sorensen:2023zkk}).

In the discussion of the EoS, an associated quantity is the sound velocity, whose behavior reflects the stiffness of the EoS and directly connects to the conformality of nuclear matter (NM). In the low-density regime, from dilute matter to compact star matter, achieving a stiff EoS that can accommodate stars with masses $\gsim 2.0M_\odot$ strongly favors exceeding the conformal limit  $c_s^2 = 1/3$ (using natural units in this work) from below~\cite{Kojo:2020krb,Brandes:2023bob,Altiparmak:2022bke,Mroczek:2023zxo}.
However, at superhigh density, one expects $c_s^2 \to 1/3$ because of the asymptotic freedom of QCD~\cite{Freedman:1977gz,Kurkela:2009gj,Fraga:2013qra}. This implies that, at intermediate density, the sound velocity is not monotonic along the density axis but instead exhibits at least one peak. Note that, after the peak, where and how the sound velocity approaches the conformal limit $c_s^2 \simeq 1/3$ is still under debate. It may or may not be at a density relevant to the cores of massive neutron star~\cite{Tews:2018kmu,Ma:2018qkg,Annala:2019puf,Ma:2020hno,Kapusta:2021ney,Margueron:2021dtx}.

The microscopic origin of the peak of sound velocity is an interesting and significant topic.
There are several debated reasons for its emergence: the transition from hadronic phase to quarkyonic phase~\cite{McLerran:2018hbz,Jeong:2019lhv,Zhao:2020dvu,Margueron:2021dtx}, topology change from baryonic to half-baryonic matter~\cite{Ma:2018xjw,Ma:2019ery,Ma:2021nuf,Lee:2021hrw,Ma:2023ugl}, the transition from  chiral symmetry breaking phase to that with a gapped Fermi surface~\cite{Hippert:2021gfs}, the hadron-quark transition~\cite{Alford:2004pf,Baym:2019iky,Chiba:2024cny,Chiba:2023ftg}, and the formation of the diquark gap ~\cite{Alford:2017qgh,Leonhardt:2019fua}. In summary, the existing arguments conclude that the peak of sound velocity emerges from the {presupposed} change in matter constituents: transition from hadronic matter to exotic configurations or emergence of new degrees of freedom (DoFs).

Recently, by analyzing the behavior of the trace anomaly in neutron star matter, Fujimoto {\it et al.} attributed the peak of sound velocity to the derivative of the trace anomaly, which steeply approaches the conformal limit and signifies the conformality~\cite{Fujimoto:2022ohj}.
This conclusion is supported by a subsequent analysis using piecewise sound velocity at different densities with respect to the constraints from massive stars~\cite{Marczenko:2022jhl}.
That is, the density dependence of sound velocity closely relates to how the scale symmetry manifests in the nuclear medium.

In this Letter, we employ a conceptually novel approach---generalized nuclear effective field theory (EFT) (GnEFT)---that extends beyond the standard chiral EFT to higher densities $n > n_0$ (with $n_0 \approx 0.16~$fm$^{-3}$ being the normal NM density). This approach, as previously developed, regards the trace anomaly of QCD as the source of the scalar meson, or dilaton, introduced through the conformal compensator. In addition to the scalar meson, the lowest-lying vector mesons $\rho$ and $\omega$ with mass $\gsim 700$~MeV are included as the relevant DoFs. Therefore, this approach is powerful and predictive up to a cutoff, $\gsim 700$~MeV, in particular for considering high densities relevant to massive compact stars~\cite{Ma:2019ery}.

We find that the peak of sound velocity naturally emerges without reference to the phase or configuration change in the matter. This novel phenomenon is attributed to the nonlinear realization of the scale symmetry, regarded as the source of scalar meson, which is different from Walecka-type quantum hadron dynamical models.  Additionally, to restore scale symmetry in dense matter, the effective $\omega$-$N$-$N$ coupling decreases faster than $\rho$-$N$-$N$ coupling in medium, thereby breaking the $\rm U(2)$ flavor symmetry, which works well in the matter-free space. These observations have not been realized before and give hints on how the scale symmetry manifests in NM, from low to high densities.

\sect{Scalar meson as a dilaton}
It has long been recognized that the trace anomaly of QCD can be taken as the source of the scalar meson field in an EFT: The scalar meson $\sigma$ is regarded as the Nambu-Goldstone (NG) boson of the spontaneous breaking of scale symmetry~\cite{Ellis:1970yd,Crewther:1970gx,Schechter:1980ak,Crewther:2013vea} triggered by its explicit breaking~\cite{Beane:1994ds}---known as the Freund-Nambu theorem~\cite{Freund:1968gyq}---which is involved in the EFT through anomaly matching.

By introducing the conformal compensator field $\chi=f_\chi\Phi=f_\chi \exp \left(\sigma / f_\chi\right)$, which under the scaling of coordinate $x_\mu \to x_\mu^\prime = \lambda^{-1}x_\mu$ transforms as $\chi(x) \to \chi^\prime(x^\prime) = \lambda \chi (x)$, and following Ref.~\cite{Crewther:2013vea}, one can write down the GnEFT~\cite{Li:2016uzn}. The mesonic Lagrangian is 
\be
\mathcal{L}_M & = & \frac{m_\rho^2}{g_\rho^2} \Phi^2 \operatorname{Tr}(\hat{\alpha}_{\|}^\mu \hat{\alpha}_{\mu \|}) \nonumber\\
& &{} + \frac{1}{2}\left(\frac{m_\omega^2}{g_\omega^2}-\frac{m_\rho^2}{g_\rho^2}\right) \Phi^2 \operatorname{Tr}(\hat{\alpha}_{\|}^\mu) \operatorname{Tr}\left(\hat{\alpha}_{\mu \|}\right) \nonumber\\
& &{} - V(\chi) + \cdots, 
\label{eq:LM}
\ee
where $\cdots$ stands for the vanished terms after taking the mean field approximation and $\hat{\alpha}_{\mu \|} = {}-\frac{1}{2}\left(g_\omega \omega_\mu+g_\rho \rho_\mu^a \tau^a\right) +\cdots$ with $g_\omega$ and $g_\rho$ being the hidden gauge coupling constants.
The explicit scale symmetry breaking effect is included in the potential,
\be
V(\chi) & = & -h_5 \Phi^4-h_6 \Phi^{4+\beta^{\prime}} ,
\ee
with $\beta^{\prime}$ being the anomalous dimension of the gluon field. With an appropriate choice of the signs of $h_5$ and $h_6$ the potential would be in the Nambu-Goldstone mode when $\beta^\prime \ne 0$. Parameters $h_5$ and $h_6$ are constrained by the saddle point equation of the dilaton. 

It has been established that this GnEFT, in what is referred to as a “leading order scale symmetry" (LOSS) approximation, is surprisingly successful, with very few parameters, for describing not only NM at the saturation density but also compact-star matter at $n \approx (5-7)n_0$~\cite{Ma:2019ery}.

Following the same reasoning, the baryonic Lagrangian is written as~\cite{Li:2016uzn} 
\be
\label{eq:LB}
\mathcal{L}_B & = & \bar{N} i \gamma_\mu D^\mu N-m_N \Phi \bar{N} N \nonumber\\
& &{} - g_{\omega NN} \omega^\mu \bar{N} \gamma_\mu N - g_{\rho NN} \rho^{a\mu} \bar{N}\tau^a \gamma_\mu N \nonumber\\
& &{} - g_{\omega NN}^{SSB} \left(\Phi^{\beta^{\prime}}-1\right) \omega^\mu \bar{N} \gamma_\mu N \nonumber\\
& &{} - g_{\rho NN}^{SSB} \left(\Phi^{\beta^{\prime}}-1\right) \rho^{a\mu} \bar{N}\tau^a \gamma_\mu N .
\ee
In Eq.~\eqref{eq:LB}, we include corrections to the LOSS that disappear in matter-free space but are found to be significant for understanding the weak decay of heavy nuclei~\cite{Ma:2020tsj}. 

\sect{Nuclear matter properties}
To saturate the constraints from symmetric nuclear matter properties around saturation density, we include the medium effect on the bare parameters in the Lagrangian using Brown-Rho scaling~\cite{Brown:1991kk,Brown:2001nh}, which is also based on the dilaton compensator approach. Explicitly, 
\be
\label{eq:br-scaling}
\frac{f_{\pi(\chi)}^*}{f_{\pi(\chi)}} & \approx & \frac{m_{\rho(\omega,N)}^*}{m_{\rho(\omega,N)}} \approx \Phi^\ast < 1~~~~ \mbox{$n \neq 0$}\ ,\nonumber\\
\frac{m_\sigma^*}{m_\sigma} & \approx &\left(\Phi^\ast\right)^{1+\frac{\beta^{\prime}}{2}}< 1~~~~ \mbox{ $n \neq 0$}\ ,
\ee
where ``\(*\)" denotes the medium modified parameter and the relations are derived in Refs.~\cite{Brown:1991kk,Li:2017hqe}.
The scaling of $h_5$ and $h_6$ can be derived from the saddle point equation and dilaton mass. In our numerical analysis, we parametrize the Brown-Rho scaling as $\Phi^\ast = 1/(1 + r n/n_0)$ with the scaling parameter $r\in\left(0.13,0.20\right)$~\cite{Paeng:2015noa}. The density dependence of the bare parameters, without loss of thermodynamic consistency~\cite{Song:1997kn,Song:2000cu}, implemented in such a way---inherited from QCD as the vacuum changes with density---is called ``intrinsic density dependence." After solving the equations of motion of hadron fields with the help of the mean field approach~\cite{Walecka:1974qa}, an induced density dependence arises from nucleon correlations, e.g., density \(\langle N^\dagger N\rangle^*\) and scalar density \(\langle \bar{N}N\rangle^*\). The full density dependence of the system is the sum of both~\cite{Ma:2019ery}.

\begin{table}[tbh]\small
    \centering
    \caption{
        Values of the saturation density $n_0$, binding energy $e_0$, symmetry energy $E_{\mathrm{sym}}$, incompressibility coefficient $K$, skewness coefficient $J$, and symmetry energy density slope $L$. $n_0$ is in units of $\mathrm{fm^{-3}}$ and others are in units of $\mathrm{MeV}$.
    }
    \begin{threeparttable}~\label{tab:nuclear-matter}
        \begin{tabular}{@{}cccc}
            \hline
            \hline
            & Empirical & bsHLS-L & bsHLS-H \\
            \hline
            $n_0$ & $0.155\pm0.050$~\cite{Sedrakian:2022ata} & 0.159 & 0.159 \\
            \hline
            $e_0$ & $-15.0\pm1.0$~\cite{Sedrakian:2022ata} & $-16.0$ & $-16.0$ \\
            \hline
            $K(n_0)$ & $230\pm30$~\cite{Dutra:2012mb} & 232 & 284\\
            \hline
            $E_{\mathrm{sym}}(n_0)$ & $30.9\pm1.9$~\cite{Lattimer:2012xj} & 30.5 & 29.2\\
            \hline
            $E_{\mathrm{sym}}(2n_0)$ & $46.9\pm10.1$~\cite{Li:2019xxz} & 51.5 & 50.2\\
            \hline
            $L(n_0)$ & $52.5\pm17.5$~\cite{Lattimer:2012xj} & 85.9 & 68.3 \\
            \hline
            $J(n_0)$ & $-700\pm500$~\cite{Farine:1997vuz} & $-767$ & $-599$\\
            \hline
            \hline
        \end{tabular}
    \end{threeparttable}
\end{table}

We find that, as shown in Table~\ref{tab:nuclear-matter}, the symmetric nuclear matter properties can be well reproduced using the parameters list in Table~\ref{tab:para}.
Although both bsHLS-L and bsHLS-H can give reasonable predictions, considering the pion-nuclei bound state data~\cite{Kienle:2004hq}, which give $r \simeq 0.2$ and the skyrmion crystal estimation~ $1.0 \lesssim |\gamma_{G^2}| \lesssim 2.0$~\cite{Ma:2016nki,Shao:2022njr}, {bsHLS-H is preferred}.
{Moreover, using bsHLS-H, we obtain $K_{\text {sym }}=9 n_0^2 \frac{\partial^2 E_{\text {sym }}}{\partial n^2}|_{n=n_0}=-54.0\ \mathrm{MeV}$ which yields \(K_\tau \approx K_{\text {sym }}-6 L(n_0)=-488\ \mathrm{MeV}\), consistent with the constraint \(K_\tau \approx-550 \pm 100~\mathrm{MeV}\)~\cite{Centelles:2008vu}. 
	}

\begin{table}[tbh]\small
    \centering
    \caption{
        Values of parameters for bsHLS-L and bsHLS-H obtained from fitting the NM properties in Table~\ref{tab:nuclear-matter}. $M_\sigma = f_\chi m_\sigma$ is in units of $10^5\rm MeV^2$.
        }
    \begin{threeparttable}~\label{tab:para}
        \begin{tabular}{@{}cccccccc}
            \hline
            \hline
            & \(M_{\sigma}\) & \(\beta'\) & \(r\) & \(g_{\omega NN}\) & \(g_{\rho NN}\) & \(g_{\omega NN}^{SSB}\) & \(g_{\rho NN}^{SSB}\)\\
            \hline
            bsHLS-L & 1.05 & 0.395 & 0.161 & 11.5 & 3.78 & 16.3 & 9.45\\
            \hline
            bsHLS-H & 2.30 & 1.15 & 0.191 & 11.0 & 4.17 & 8.85 & 4.85\\
            \hline
            \hline
        \end{tabular}
    \end{threeparttable}
\end{table}

\begin{figure}[tbh]
    \centering
    \subfigure{
        \includegraphics[width=0.455\textwidth]{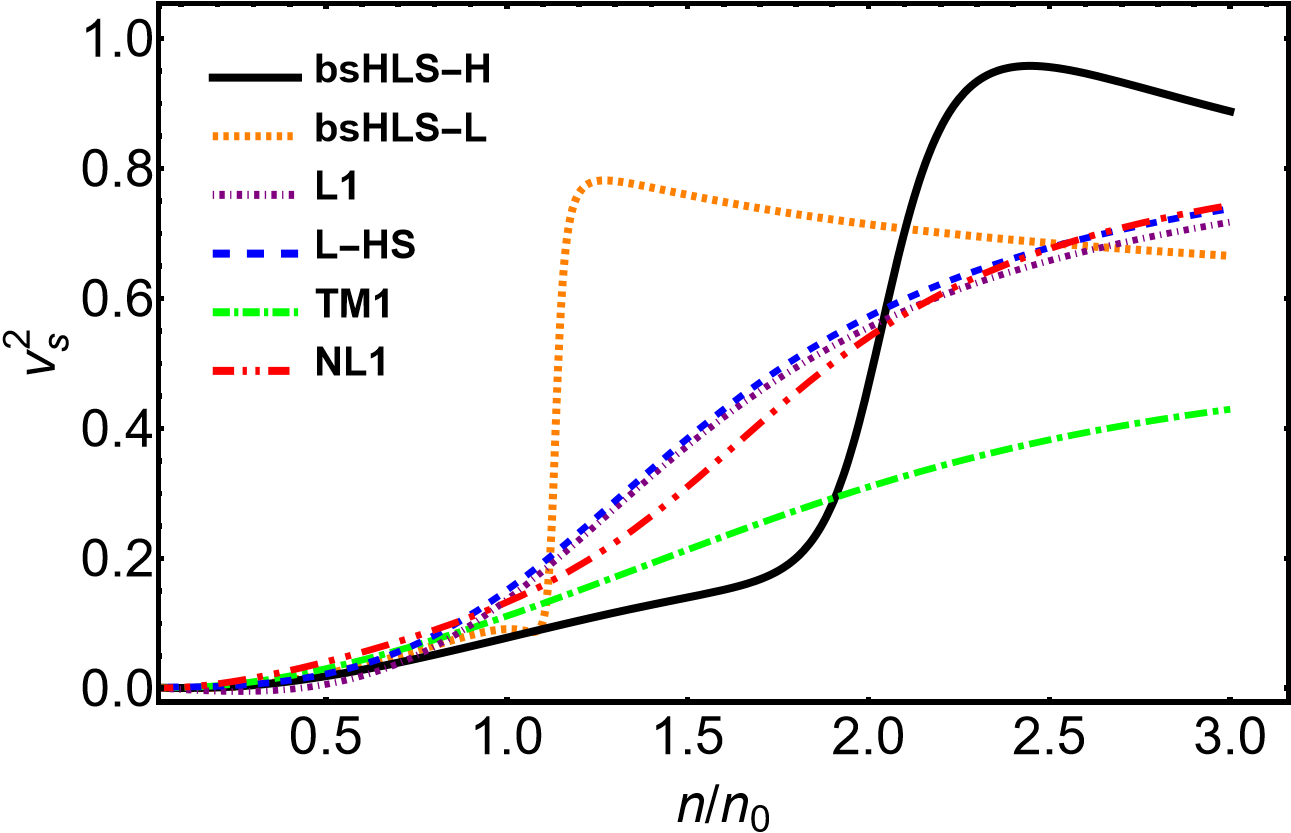}
    }
    \caption{The speed of sound in {PNM} for various models. The results from the choices of parameters in  Walecka-type models L1~\cite{Lee:1986bi}, L-HS~\cite{Horowitz:1981xw}, NL1~\cite{Reinhard:1986qq} and TM1~\cite{Sugahara:1993wz} are shown for comparison. }~\label{fig:sound-bsHLS}
\end{figure}

\sect{Speed of sound}
The most interesting observation in this work is the existence of a $v_s$ peak in pure neutron matter (PNM) which is more relevant to neutron stars at intermediate densities around $(1 - 2.5)\ n_0$ shown in Fig.~\ref{fig:sound-bsHLS}, where the $v_s$ surges at a density then goes downwards. In the studies cited above, this peak is always coming from the hadron-quark transition or configuration change. But here, only hadron DoFs exist and no new configuration emerges. In addition, this peak does not exists in the Walecka-type models~\cite{Lee:1986bi,Horowitz:1981xw, Reinhard:1986qq, Sugahara:1993wz} compared in Fig.~\ref{fig:sound-bsHLS}. Therefore, this unique feature, which not been observed in other hadronic models, emerges from the realization of the trace anomaly of QCD at the hadronic level, as will be discussed in detail later.

\begin{figure}[tbh]
    \centering
        \includegraphics[width=0.455\textwidth]{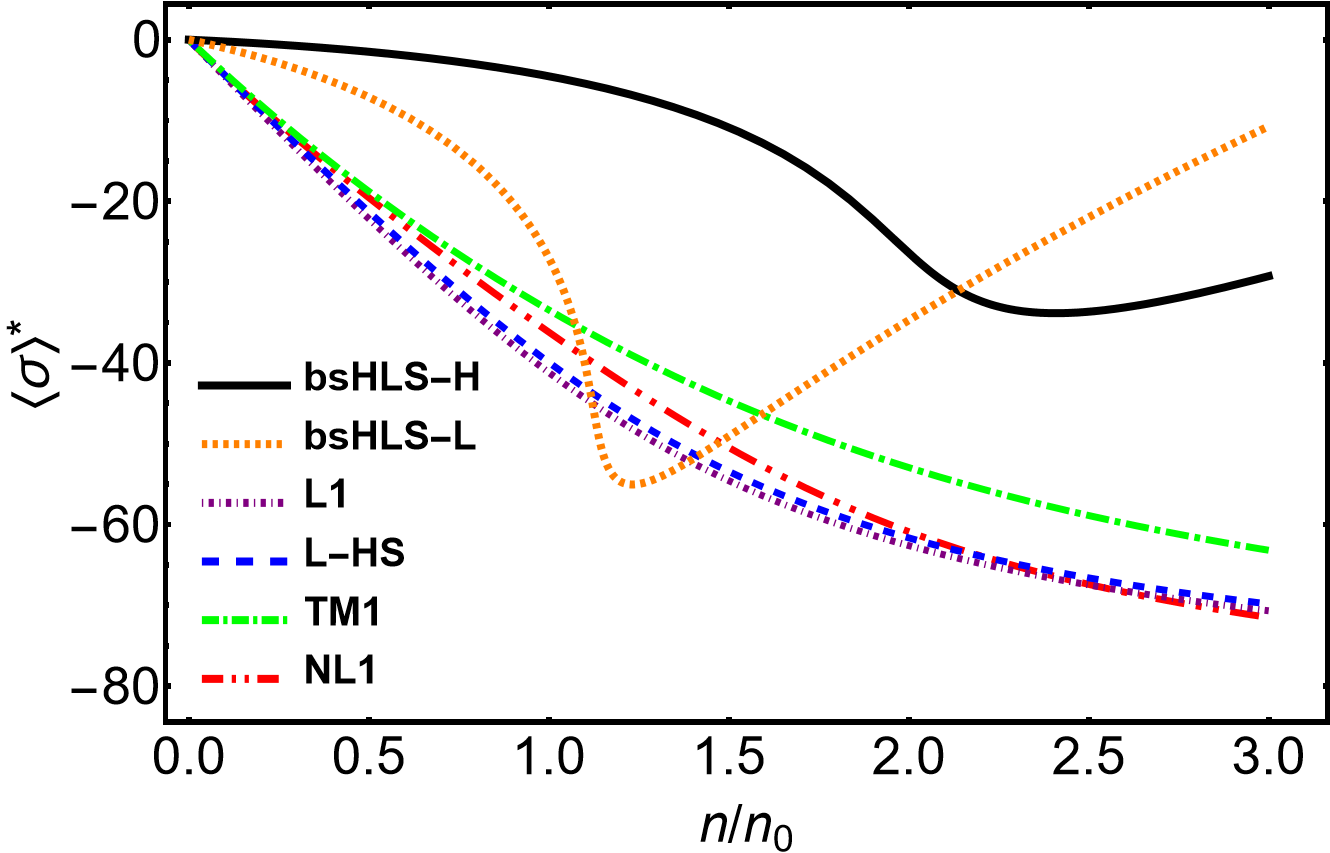}
    \caption{
        The mean field of the $\sigma$ field in {PNM}.
        \(f_\chi\) is chosen as \(f_\chi\approx f_\pi=92.4\ {\rm MeV}\) for consistency with the chiral expansion~\cite{Crewther:2013vea}.
        Notations are the same as in Fig.~\ref{fig:sound-bsHLS}.
    }~\label{fig:sigma}
\end{figure}

To have a deeper insight into the mechanism of the peak of sound velocity in the dilaton compensator approach, we compare the mean field $\sigma$ in the dilaton compensator $\chi = f_\chi e^{\sigma/f_\chi}$ and that in the widely used Walecka-type models~\cite{Lee:1986bi,Horowitz:1981xw, Reinhard:1986qq, Sugahara:1993wz} in Fig.~\ref{fig:sigma}.
It can be seen that all the linear models have a monotonic \(\sigma\) density dependence, but there are kink behaviors in both bsHLS-L and bsHLS-H. One can conclude that the peak in $v_s^2$ is due to the kink behavior of the $\sigma$ field in bsHLS-L and bsHLS-H (\(\sigma\) is nonlinearly coupled with other mesons through the conformal compensator), which does not exist in Walecka-type models.
This is the key factor that allows the bsHLS $\sigma$ field to generate the $v_s$ peak without resorting to any transition of phase or configuration.

\sect{Scale symmetry in medium}
Upon closer inspection of Fig.~\ref{fig:sigma}, one may observe that the magnitude of spontaneous scale symmetry breaking increases with density after the kink, contrary to naive expectations, for order parameter of scale symmetry: \(\langle\chi\rangle^*\) increasing with \(\langle\sigma\rangle^*\).
Actually, this phenomena is also found in the skyrmion crystal approach to dense matter using the dilaton compensated EFT~\cite{Park:2003sd,Park:2008zg,Ma:2013ela,Shao:2022njr}.
One of the reasons is that, due to the dilaton compensator, the vector meson mass scales with density, e.g., $m_\omega^\ast = \Phi m_\omega$.
Then, if the omega mass were to decrease as required by the vector manifestation~\cite{Harada:2000kb}, the repulsive omega force would increase with density. This forces the omega meson mass $m_\omega^\ast$, hence $\Phi$, to increase. The magnitude of the scale symmetry breaking increases~\cite{Park:2008zg}.

\begin{figure}[tbh]
	\centering
	\includegraphics[width=0.455\textwidth]{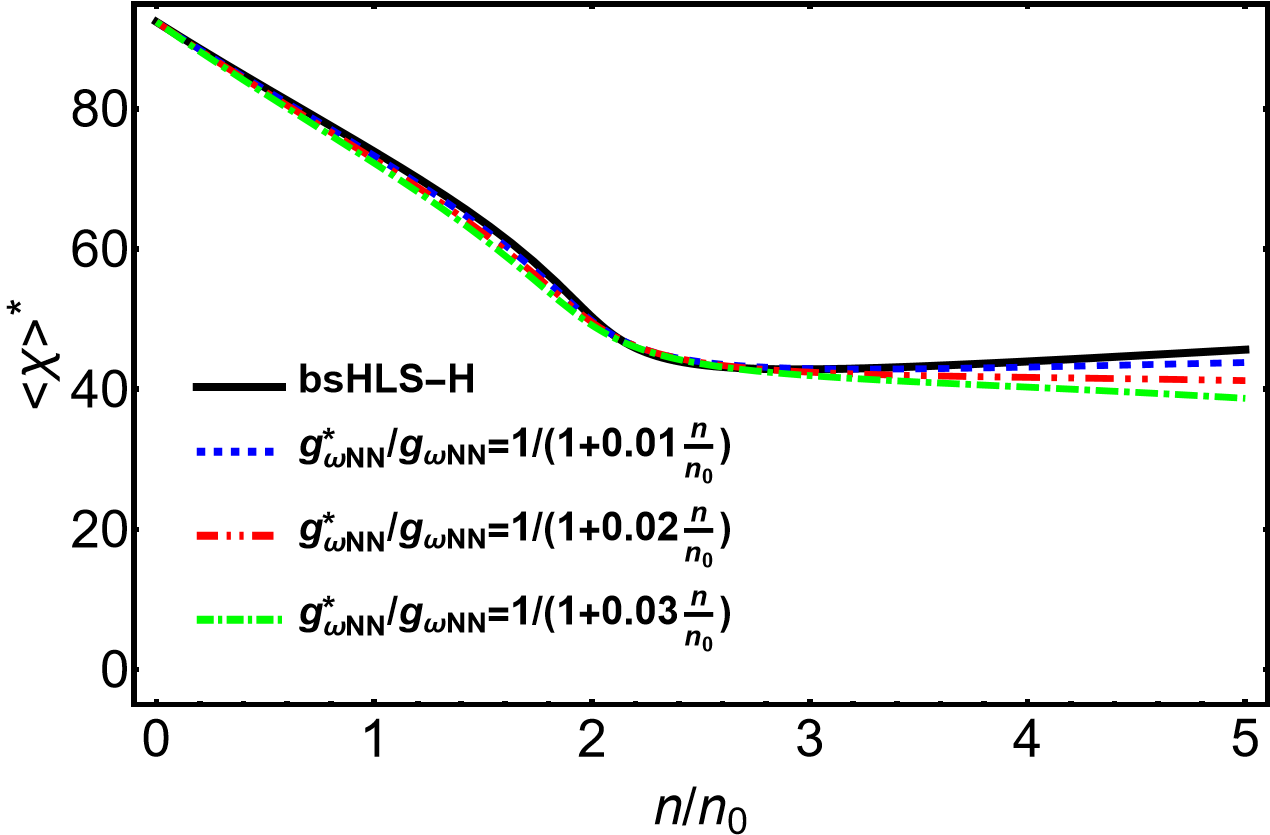}
	\caption{
		The expectation value of $\chi$ as a function of density with modified $\omega$-$N$-$N$ coupling, where \(f_{\chi}\) is set be equal to \(f_{\pi}\).
	}~\label{fig:VEVChieff}
\end{figure}

\begin{figure}[tbh]
	\centering
	\includegraphics[width=0.455\textwidth]{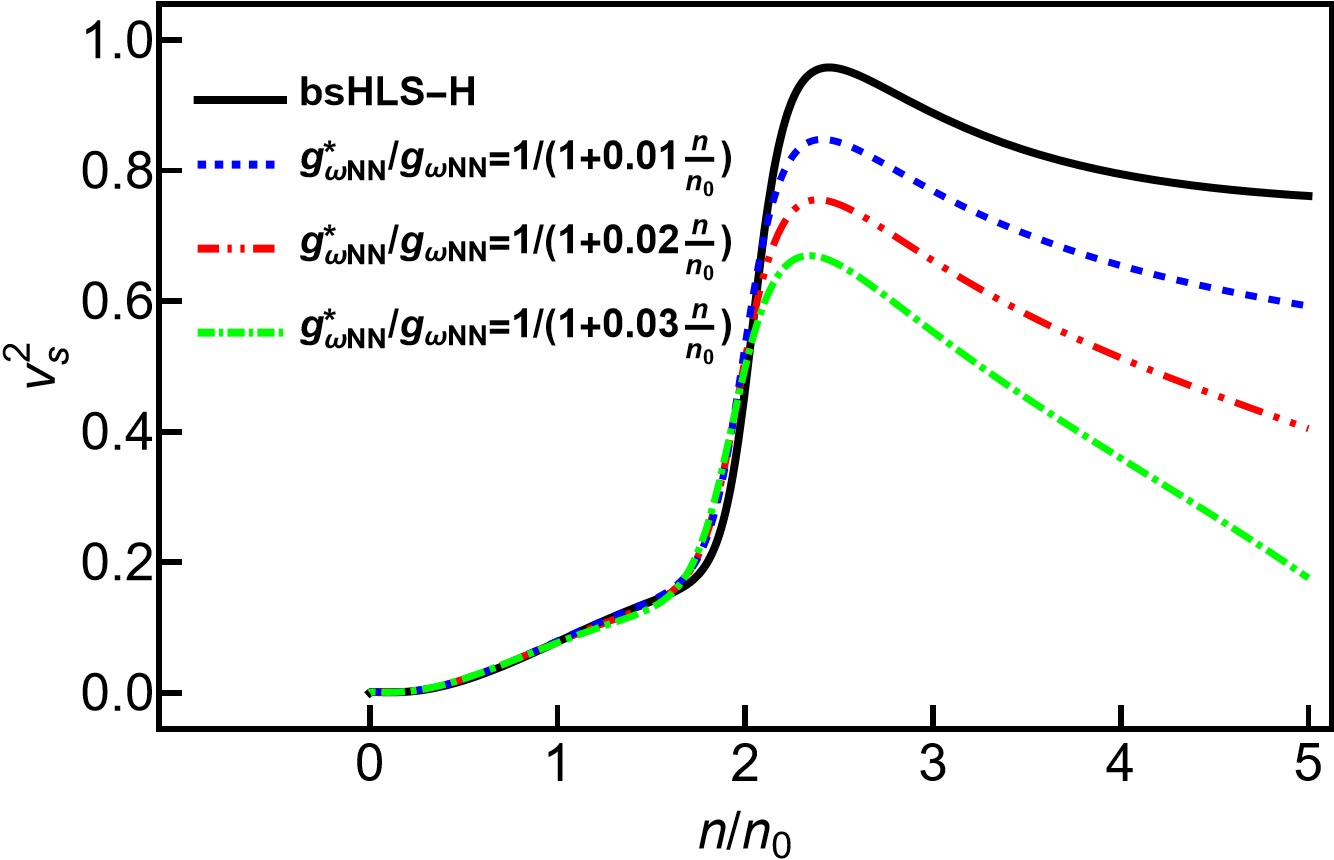}
	\caption{
		The sound velocity in {PNM} with modified $\omega$-$N$-$N$ coupling.
	}~\label{fig:sound velocityeff}
\end{figure}

One of the solutions of this flaw is to couple more dilaton fields to the $\omega$-$N$-$N$ coupling to impose stronger density scaling~\cite{Park:2008zg,Paeng:2013xya,Shao:2022njr}.
Here, we typically choose factor $g_{\omega NN}^\ast/g_{\omega NN} \equiv 1/(1+ Rn/n_0)$ and plot the vacuum expectation value (VEV) of dilaton field $\chi$ in Fig.~\ref{fig:VEVChieff}. It clearly shows that, with this further suppression, the magnitude of scale symmetry breaking after the kink is reduced.
One may find that this suppression, especially the near independence of $\langle \chi \rangle^\ast$ on density, is very similar to the pseudoconformality based on the topology consideration~\cite{Ma:2018xjw, Ma:2019ery, Ma:2021nuf, Lee:2021hrw, Ma:2023ugl}. At this moment, we cannot clearly build a connection between these two approaches. One possible way might be to look at the baryon number density distribution in the dilaton compensated chiral-quark model~\cite{MantovaniSarti:2013aqe} or  the baryon charge distribution using the Kohn-Sham theory~\cite{Hu:2019}.
An interesting observation is that the peak of sound velocity remains present, and its location is closely tied to the kink of $\langle \chi\rangle^\ast$, as illustrated in Fig.~\ref{fig:sound velocityeff}. One may find that the sound velocity in Fig.~\ref{fig:sound velocityeff} does not consistently reach the asymtotic limit of QCD if our results are naively extrapolated to higher density. This is understandable since our approach only includes hadronic DoFs and such extrapolation is not solid.

It should be noted that the significance of the corrections to LOSS has been found in the weak decays of heavy nuclei~\cite{Ma:2020tsj}.
To understand the data of the weak decay of heavy nuclei with $A > 60$~\cite{Lubos:2019nik}, a correction to the LOSS should be implemented to the weak current of the nucleon.
Similarly to Ref.~\cite{Ma:2020tsj}, the present observation illustrates how hidden scale symmetry emerges through strong nuclear correlations with an infrared (IR) fixed point realized—in the chiral limit—in the NG mode: that is, the scale symmetry manifests in the nuclear medium, for continuity from the unitarity limit at low density (in light nuclei) to the dilaton limit at high density (in compact stars).
However, between these limits, such as at normal NM density, the symmetry is not visible, hence hidden.

\sect{Conclusion and remarks}
In this work, we clearly elucidated that the peak of sound velocity in NM is locked to how the scale symmetry is implemented in quantum hadron dynamics. 

In stark contract to what was proposed in the existing literature, that this peak emerges due to the presupposed transition from a phase with broken chiral symmetry to one with quark or exotic configurations, we found that it emerges naturally after solving the coupled equations of motion of hadron fields in the mean field approximation in a pure hadronic EFT with broken chiral symmetry and the scalar meson introduced as a dilaton---the NG boson of the broken scale symmetry---via the conformal compensator approach.

Explicit calculations show that this peak does not emerge in Walecka-type models, where the sigma meson field is involved without strong symmetry motivation. In the dilaton compensator approach considered here, the dilaton---the scalar meson---couples to the hadron mass terms, which does not happen in the Walecka-type models. Therefore, the peak of sound velocity in this work is intrinsic to the pattern of scale symmetry in nuclear matter and the character of the lightest scalar meson, whether it is an NG boson of the broken scale symmetry of QCD or the fourth component of the chiral four-vector. 

Due to the constraints from symmetric nuclear matter properties around saturation density, the location of the peak of sound velocity of PNM is estimated to be $(1 - 2.5)\ n_0$ which is a promising range for terrestrial experiments, such as heavy ion collisions at fscilities HIAF, RAON, RHIC, FRIB, and so on~\cite{Sorensen:2023zkk}. In addition, it is found that the radius of the typical neutron star with a mass of $1.4~M_\odot$ is $\approx 13.2$~km for the bsHLS-H set, which is within the constraint of GW170817~\cite{LIGOScientific:2018cki,LIGOScientific:2018hze} and validates the present approach. Moreover, the maximum neutron star mass is predicted to be $2.85\ M_{\odot}$, larger than the existing results from pure hadronic models~\cite{Ozel:2016oaf, Guo:2023mhf}, which serves as a future test. A detailed analysis of the NS properties will be given in a forthcoming publication. Of course we admit that, while our model's predictions are aligned with all the known empirical data, we connot definitely distinguish it from the others in the literature at this moment. The merit of this work is tat it provides a mechanism for the origin of the peak of sound velocity, which has not been explicitly pointed out before.

In addition to the information on the manifestation of scale symmetry in hadronic matter focused on in this work, the observation discussed above is also valuable for understanding the mechanism of electroweak symmetry breaking~\cite{Goldberger:2007zk,Matsuzaki:2013eva} and the evolution of our universe~\cite{Garcia-Bellido:2011kqb}.

\sect{Acknowledgments}
The work of Y.~L. M. is supported in part by the National Science Foundation of China (NSFC) under Grant No. 12347103, the National Key R\&D Program of China under Grant No. 2021YFC2202900 and Gusu Talent Innovation Program under Grant No. ZXL2024363.

\sect{Data availability}
No data were created or analyzed in this study.

	

\bibliography{RefPeakSVScale}

\end{document}